\documentclass[12pt]{article}

\usepackage{amsmath}
\usepackage{amsfonts}
\usepackage{amssymb}
\usepackage{cite}
\usepackage{graphics,graphicx,tikz}
\usepackage{soul}
\usetikzlibrary{patterns}

\numberwithin{equation}{section} 
\usepackage[linktocpage]{hyperref}
\hypersetup{
	colorlinks=true,
	linkcolor=blue,
	filecolor=magenta,      
	urlcolor=blue,
	citecolor=blue
}
\def\beq{\begin{equation}}
\def\eeq{\end{equation}}
\def\bea{\begin{align}}
\def\eea{\end{align}}

\usepackage{color}

\begin{document}
\begin{titlepage}
\hfill \hbox{CERN-TH-2021-008}
\vskip 0.1cm
\hfill \hbox{NORDITA 2021-001}
\vskip 0.1cm
\hfill \hbox{QMUL-PH-21-03}
\vskip 0.1cm
\hfill \hbox{UUITP-03/21}
\vskip 1.0cm
\begin{flushright}
\end{flushright}
\vskip 1.0cm
\begin{center}
{\Large \bf Radiation Reaction from Soft Theorems}
\vskip 1.0cm {\large  Paolo Di Vecchia$^{a, b}$, Carlo Heissenberg$^{b,c}$,
Rodolfo Russo$^{d}$, \\
Gabriele Veneziano$^{e, f}$ } \\[0.7cm]

{\it $^a$ The Niels Bohr Institute, Blegdamsvej 17, DK-2100 Copenhagen, Denmark}\\
{\it $^b$ NORDITA, KTH Royal Institute of Technology and Stockholm University, \\
 Roslagstullsbacken 23, SE-10691 Stockholm, Sweden}\\
 {\it $^c$ Department of Physics and Astronomy, Uppsala University,\\ Box 516, SE-75120 Uppsala, Sweden}\\
{\it $^d$ Queen Mary University of London, Mile End Road,\\ E1 4NS London, United Kingdom}\\
{\it $^e$ Theory Department, CERN, CH-1211 Geneva 23, Switzerland}\\
{\it $^f$Coll\`ege de France, 11 place M. Berthelot, 75005 Paris, France}
\end{center}
\begin{abstract}
Radiation reaction (RR) terms at the third post-Minkowskian (3PM) order have recently been found to be instrumental in restoring smooth continuity between the non-relativistic, relativistic, and ultra-relativistic (including the massless) regimes. Here we propose a new and intriguing connection between RR and soft (bremsstrahlung) theorems which short-circuits the more involved conventional loop computations. Although first noticed in the context of the maximally supersymmetric theory,  unitarity and analyticity arguments support the general validity of this 3PM-order connection that we apply, in particular, to Einstein's gravity and to its Jordan-Brans-Dicke extension. In the former case we find full agreement with a recent result by Damour obtained through a very different reasoning.
\end{abstract}
\end{titlepage}


\section{Introduction}
\label{sec:intro}

The gravitational scattering of classical objects at large impact parameter $b$ is relevant for the study of the inspiral phase of black-hole binaries since it can be used to determine the parameters of the Effective-One-Body description (see~\cite{Damour:2016gwp} and references therein). For this reason, gravitational scattering has been at the centre of renewed attention and has been recently investigated using a variety of techniques, including the use of quantum field theory (QFT) amplitudes to extract the relevant classical physics~\cite{Goldberger:2004jt,Melville:2013qca,Goldberger:2016iau,Luna:2016idw,Luna:2017dtq,Bjerrum-Bohr:2018xdl,Cheung:2018wkq,Kosower:2018adc,Bern:2019nnu,KoemansCollado:2019ggb,Cristofoli:2019neg,Bern:2019crd,Kalin:2019rwq,Bjerrum-Bohr:2019kec,Kalin:2019inp,Damour:2019lcq,Cristofoli:2020uzm,Kalin:2020mvi,Kalin:2020lmz,Kalin:2020fhe,Mogull:2020sak,Huber:2020xny}. Here we will focus in particular on the eikonal approach~\cite{Amati:1987wq, Amati:1987uf,Muzinich:1987in, Sundborg:1988tb}, where the classical gravitational dynamics is derived from standard QFT amplitudes by focusing on the terms that exponentiate in the eikonal phase $e^{2i \delta}$. The Post-Minkowskian (PM) expansions writes $\delta$ as a perturbative series in the Newton constant $G$ at large values of $b$ and the state-of-the-art results determine the real part of the 3PM ({\em i.e.} 2-loop) eikonal $\operatorname{Re} 2\delta_2$ (or the closely related scattering angle) and to some extent the imaginary part, both in standard GR~\cite{Amati:1990xe,Bern:2019nnu,Bern:2019crd,Cheung:2020gyp,Kalin:2020fhe,Damour:2020tta} and various supersymmetric generalisations~\cite{DiVecchia:2019kta,Bern:2020gjj,Parra-Martinez:2020dzs,DiVecchia:2020ymx}.

In this letter we expand on the approach discussed in~\cite{Amati:1990xe,DiVecchia:2020ymx} where the relation between the real and the imaginary part of $\delta_2$ was used to derive the 3PM scattering angle in the ultrarelativistic limit and to show that it is a universal feature of all gravitational theories in the two derivative approximation. 
Furthermore, it was shown in~\cite{DiVecchia:2020ymx} for ${\cal N}=8$ supergravity that taking into account the full soft region in the loop integrals was crucial to obtain a smooth interpolation between the behaviour of $\delta_2$ in the non-relativistic, {\em i.e.} Post-Newtonian (PN), regime and the  ultrarelativistic (or massless) one. The additional contributions coming from the full soft region had the feature of contributing half-integer terms in the PN expansion and were therefore interpreted as radiation-reaction (RR) contributions. This connection was further confirmed in~\cite{Damour:2020tta} by Damour, who used a linear response relation earlier derived in~\cite{Bini:2012ji} to connect these new RR terms to the loss of angular momentum in the collision. In this way the result of~\cite{DiVecchia:2020ymx} was extended to the case of General Relativity~\cite{Damour:2020tta}.

In this paper we argue that there is actually a  direct  relation between the RR and the much studied soft-bremsstrahlung limits. We claim that the real part of the RR eikonal at 3PM (indicated by ${\rm Re}\, 2 \delta_2^{(rr)}$)  is simply related to the infrared divergent contribution of its imaginary part $({\rm Im}\, 2 \delta_2)$. This relation holds at all energies and reads
\begin{equation}
\lim_{\epsilon\to 0} {\rm Re} \,2 \delta_2^{(rr)} = -\lim_{\epsilon\to 0}\left[ \pi \epsilon ({\rm Im}\,2 \delta_2)\right]
\;, \label{1.5}
\end{equation}
where, as usual, $\epsilon = \frac{4-D}{2}$ is the dimensional regularisation parameter. On the other hand, there is a  simple connection (see e.g.~\cite{Addazi:2019mjh}) between the infrared divergent  imaginary part of $\delta_2$ and the so-called zero-frequency limit~\cite{Smarr:1977fy} of the bremsstrahlung spectrum reading: 
\begin{equation}
  \lim_{\epsilon\to 0} \left[ - 2\epsilon ({\rm Im}\,2 \delta_2)\right] = \frac{d E^{rad}}{2 \hbar d \omega}(\omega \to 0) ~ \Rightarrow~ 
  \lim_{\epsilon\to 0}{\rm Re} \,2 \delta_2^{(rr)} = \frac{\pi}{4 \hbar} \frac{d E^{rad}}{d \omega}(\omega \to 0)
\;, \label{ZFL}
\end{equation}
so that, in the end, RR gets directly related to soft bremsstrahlung. We stress that all (massless) particles can contribute to the r.h.s. of \eqref{ZFL} and therefore to the RR. 

This result was first noticed in the ${\cal N}=8$ supergravity setup of~\cite{Caron-Huot:2018ape,Parra-Martinez:2020dzs} by using the results of~\cite{DiVecchia:2020ymx,DiVecchia:2021bdo}, see also~\cite{Herrmann:2021tct}, where the full 3PM eikonal is derived by a direct computation of the 2-loop amplitude describing the scattering of two supersymmetric massive particle. Here we give an interpretation of this connection and conjecture its general validity in gravity theories at the 3PM level  (the first non trivial one) by reconstructing the infrared divergent part of ${\rm Im} \, 2 \delta_2$ from the three-body discontinuity involving the two massive particles and a massless particle. The building block is of course the $2 \rightarrow  3$ five-point tree-level amplitude where, for our purposes, it is sufficient to keep only the leading classical divergent term in the soft limit (the so-called Weinberg term) of the massless particle.      

When focusing on pure GR, the only massless particle that can be involved in the three-particle cut mentioned above is the graviton. We will see that, by using Eq. \eqref{1.5}, we reproduce  the deflection angle recently derived in~\cite{Damour:2020tta} on the basis of a linear-response formula and of a lowest-order calculation of the angular momentum flux. In the  massive ${\cal N}=8$ case, one needs to consider, in addition to the graviton, the contributions of the relevant vectors and scalar fields (including the dilaton). Once all massless particles that can appear in the three-particle cut are taken into account, one obtains~\eqref{3.6} which, as already mentioned, satisfies~\eqref{1.5}.
The basic idea underlying all cases is that the calculation of ${\rm Im}\, 2 \delta_2$ from sewing tree-level, on shell, inelastic amplitudes is far simpler than the derivation of the full two-loop elastic amplitude even when focusing on just  the classical contributions. Both for GR and for $\mathcal N=8$, the infrared divergent piece of $\delta_2$ can be equivalently obtained exploiting the exponentiation of infrared divergences in momentum space for the elastic amplitude itself (details will be presented elsewhere). The arguments supporting~\eqref{1.5} appear to be valid within a large class of gravitational theories and so this equation provides a direct, general way to calculate the RR contributions at the 3PM level. It remains to be seen whether this approach can be generalized, and in which form, beyond 3PM.

The paper is organized as follows. In Sect.~\ref{softm} we introduce our kinematical set-up for the relevant elastic ($2 \rightarrow 2$) and inelastic ($2 \rightarrow 3$) processes and discuss the standard soft limit of the latter in momentum space. In Sect.~\ref{RRIR} we present the empirical connection between $\operatorname{Re} 2\delta_2^{(rr)}$  and the IR divergent  part of $\operatorname{Im} 2\delta_2$ in the maximally supersymmetric case. Using unitarity and  analyticity  of  the scattering amplitude, we provide arguments in favour of its general validity. We also outline the logic of the calculations that follow.  In Sect.~\ref{softb} we transform the soft-limit results of Sect.~\ref{softm} to impact-parameter space in the large-$b$ limit. In Sect.~\ref{probability} we use these to compute the divergent part of $\operatorname{Im} 2 \delta_2$ and, through our connection, the RR terms in $\operatorname{Re} 2  \delta_2$. This is first done for the case of ${\cal N}=8$ supergravity, where we recover the result of~\cite{DiVecchia:2020ymx}, and then for Einstein's gravity, reproducing the result of~\cite{Damour:2020tta}, and for Jordan-Brans-Dicke theory.
  
\section{Soft Amplitudes in Momentum Space}
\label{softm}

Let us start by better defining the processes under consideration. We shall be interested in the scattering of two massive scalar particles in $D=4-2\epsilon$ dimensions, with or without the additional emission of a soft massless quantum. For GR, we thus consider minimally coupled scalars with masses $m_1$, $m_2$ in $4-2\epsilon$ dimensions. For $\mathcal N=8$ supergravity, that can be obtained by compactifying six directions in ten-dimensional type II supergravity,  we instead choose incoming Kaluza--Klein (KK) scalars whose $(10-2\epsilon)$-dimensional momenta read as follows:
  \begin{equation}
    \label{eq:kin10D}
     P_1= (p_1;0,0,0,0,0,m_1)\,,\qquad P_2 = (p_2;0,0,0,0,m_2 \sin\phi,m_2 \cos\phi)\;,
   \end{equation}
where the last six entries refer to the compact KK directions and provide $p_1$, $p_2$ with the desired effective masses $m_1$, $m_2$ in $4-2\epsilon$ dimensions. The  angle $\phi$ thus describes the relative orientation between the KK momenta,

We work in a centre-of-mass frame and for our purposes it is convenient to regard the amplitudes as functions of $\bar{p}$, encoding the classical momentum of the massive particles, the transferred momentum $q$ (which is related to the impact parameter after Fourier transform) and the emitted momentum $k$. We thus parametrise the momenta of the incoming states as follows,
\begin{equation}
    \label{eq:kin}
    \begin{aligned}
    p_1&= (E_1,\vec{p}\,) = \bar{p}_1 - a q + c k \,,
     &&\bar{p}_1= (E_1,0,\ldots,0, \bar{p}\,) \,,
     \\
     p_2&= (E_2,-\vec{p}\,) = \bar{p}_2 + a q - c k \,,
     &&\bar{p}_2= (E_2,0,\ldots,0, -\bar{p}\,) \,,
    \end{aligned}
\end{equation}
while the outgoing\footnote{We treat all vectors as formally ingoing.} states are a soft particle of momentum $k$ and massive states with momenta
\begin{equation}
    \label{eq:kin2}
    k_1= -\bar{p}_1 - (1-a) q - c k \,,\qquad \ \; \,
    k_2=  -\bar{p}_2 + (1-a) q - (1-c) k \,.
\end{equation}
We singled out the direction of the classical momentum $\bar{p}$, while $q$ is non-trivial only along the $2-2\epsilon$ space directions orthogonal to $\bar{p}_i$. In the elastic case of course $k=0=c$ and we have $a=1/2$. For the inelastic amplitudes one can fix $a$ and $c$ by imposing the on-shell conditions and using $\bar{p}_i q=0$, but we will not need their explicit expression in what follows. 

We shall now collect the tree-level amplitudes that will enter our calculation of $\operatorname{Im}\, 2\delta_2$ via unitarity, focusing for the most part on $\mathcal N=8$ and commenting along the way on small amendments that are needed to obtain the GR amplitudes.
	
The simplest building block for our analysis of $\mathcal N=8$ supergravity is the elastic tree-level amplitude
\begin{equation}
A_{tree} \simeq - \frac{32\pi G  m_1^2 m_2^2 (\sigma - \cos \phi)^2}{t}\,,\qquad\text{with } \sigma = - \frac{p_1p_2}{m_1 m_2}\,,
\label{2.1} 
\end{equation}
where we retained only the terms with the pole at $t=-q^2 = 0$, since we restrict our attention to  long-range effects. When $\phi=\frac{\pi}{2}$, the KK momenta are along orthogonal directions, and, in this case,  the pole at $t = 0$ corresponds to the exchange of the graviton and of the dilaton that are coupled universally to all massive states with the following three-point on-shell amplitudes in $D=4$:
\begin{equation}
A_3^{\mu \nu} =-i \kappa ( p_j^\mu k_j^\nu + p_j^\nu k_j^\mu )\,,\qquad
A_3^{dil} = -i \kappa \sqrt{2}\, m_j^2\;,
\label{2.2}
\end{equation}
with $j=1,2$ and $\kappa = \sqrt{8 \pi G}$. Using the vertices~\eqref{2.2} and standard propagators, the graviton and the dilaton exchanges yield
\begin{equation}
A^{gr}_{tree} \simeq - \frac{16\pi G  m_1^2 m_2^2 (2 \sigma^2 - 1)}{t} ~,\quad A^{dil}_{tree} \simeq -\frac{16\pi G  m_1^2 m_2^2 }{t} ~.
\label{2.1b} 
\end{equation}
Their sum reproduces~\eqref{2.1} for $\phi=\frac{\pi}{2}$. For generic $\phi$, in addition to the couplings mentioned above, we also need to consider massless vectors and scalars coming from the KK compactification of the ten dimensional graviton. We have a scalar and a vector whose three-point amplitudes involving the massive fields are
\begin{equation}
\begin{aligned}
  A_3^\mu &= - i \kappa m_1 \sqrt{2} (p_1-k_1)^\mu \,, \qquad \qquad \; A_3 = - i \kappa 2 m_1^2  \;,  \\
  A_3^\mu &= - i \kappa m_2 \sqrt{2} (p_2-k_2)^\mu \cos\phi\,, \qquad A_3 = - i \kappa 2 m_2^2 \cos^2 \phi \;.
\end{aligned}
\label{2.3}
\end{equation}
Including also the contribution of these states one can reproduce the tree-level amplitude~\eqref{2.1} for $\phi=0$, which provides a useful cross-check for the normalization of the three-point amplitudes. The particle with mass $m_2$ couples to another vector and another scalar with a strength depending to the other component of the KK momentum,
\begin{equation}
  B_3^\mu = - i \kappa m_2 \sqrt{2} (p_2-k_2)^\mu \sin\phi\,,\qquad  B_3 = - i \kappa 2 m_2^2 \sin^2 \phi \,.
  \label{2.3b}
  \end{equation}
There is also an extra scalar related to the off-diagonal components of the internal metric whose coupling is proportional to $\cos\phi\sin\phi$; here we will not use this coupling as we will mainly focus on the cases $\phi=0$ and $\phi=\frac{\pi}{2}$. 

Let us now move to the inelastic, $2 \to 3$ amplitude. As stressed in the introduction, we can restrict ourselves  to the leading  soft term that diverges as $k ^{-1}$ for $k\to 0$. It is given by the product of the elastic tree-level amplitude  times a soft factor.
For instance, the leading term for the emission of a soft graviton is \cite{Weinberg:1965nx}:
\begin{equation}
  \label{eq:grnde}
  A_5^{\mu\nu} \simeq {\kappa} \left(\frac{p_1^\mu p_1^\nu}{p_1 k} + \frac{k_1^\mu k_1^\nu}{k_1 k} + \frac{p_2^\mu p_2^\nu}{p_2 k} + \frac{k_2^\mu k_2^\nu}{k_2 k} \right) A_{tree}\,,
\end{equation}
while in the case of the dilaton one finds\footnote{We neglect possible terms proportional to $\delta(\omega)$ which play no role in the present discussion.} \cite{DiVecchia:2015jaq}
\begin{equation}
  \label{eq:dinde}
  A_5^{dil} \simeq -\frac{\kappa}{\sqrt{2}} \left(\frac{m_1^2}{p_1 k} + \frac{m_1^2}{k_1 k} + \frac{m_2^2}{p_2 k} + \frac{m_2^2}{k_2 k} \right) A_{tree}\,.
\end{equation}

We now use~\eqref{eq:kin} and~\eqref{eq:kin2} and keep the leading terms in the soft limit $k\to 0$. By further keeping only the {\em classical} contributions, which are captured by the linear terms in the $q \to 0$ limit, one obtains
\begin{equation}
A_5^{\mu \nu} \simeq  \kappa \left[  \left(\frac{\bar{p}_1^\mu \bar{p}_1^\nu}{(\bar{p}_1k)^2} - \frac{\bar{p}_2^\mu \bar{p}_2^\nu}{(\bar{p}_2k)^2} \right) (qk) -  \frac{\bar{p}_1^\mu q^\nu+ \bar{p}_1^\nu q^\mu}{(\bar{p}_1k)  }   + \frac{\bar{p}_2^\mu q^\nu+ \bar{p}_2^\nu q^\mu}{(\bar{p}_2k) }  \right] A_{tree}
\label{2.4}
\end{equation}
for the graviton
and
\begin{equation}
  \label{eq:dfe}
  A_5^{dil} \simeq -\frac{\kappa}{\sqrt{2}} \left( \frac{m_1^2 (qk)}{(\bar{p}_1k)^2} -  \frac{m_2^2 (qk)}{(\bar{p}_2k)^2} \right)A_{tree} 
\end{equation}
for the dilaton.

  From now on we focus for simplicity on the case $\phi=\frac{\pi}{2}$ and so only the first line of~\eqref{2.3} is non-trivial; together with the contribution of~\eqref{2.3b} we need to consider the emission of the two vectors and of the two scalars. For the soft amplitudes we find:
\begin{equation}
A_5^\mu \simeq  \kappa m_1 \sqrt{2} \left( \frac{\bar{p}_1^\mu (qk)}{(\bar{p}_1k)^2} - \frac{q^\mu}{\bar{p}_1k}\right) A_{tree} \,, \quad B_5^\mu \simeq \kappa m_2 \sqrt{2} \left(- \frac{\bar{p}_2^\mu (qk)}{(\bar{p}_2k)^2} + \frac{q^\mu}{\bar{p}_2k}\right)A_{tree} \; ,
\label{2.5}
\end{equation}
\begin{equation}
A_5 \simeq \kappa m_1^2 \frac{(qk)}{(\bar{p}_1k)^2} A_{tree} \,,\qquad B_5 \simeq - \kappa m_2^2 \frac{(qk)}{(\bar{p}_2k)^2} A_{tree} \;.
\label{2.6}
\end{equation}

\section{Radiation Reaction from Infrared Singularities}
\label{RRIR}

In this section we briefly present our arguments for the validity, at two-loop level and for generic gravity theories,  of the relation \eqref{1.5}. We leave a more detailed discussion to a longer paper \cite{DiVecchia:2021bdo}.

Our starting point is an empirical observation made in the context of a recent calculation in ${\cal{N}}=8$ supergravity \cite{DiVecchia:2020ymx} whose set-up has been recalled in the previous section. An interesting outcome of that calculation (made for $\cos \phi =0$) was the identification of a radiation-reaction contribution  to the real part of the (two loop) eikonal phase,  given by 
\begin{equation}
{\rm Re}\,2 \delta_2^{(rr)}  = \frac{16 G^3 m_1^2 m_2^2 \sigma^4}{\hbar b^2 (\sigma^2-1)^2} \Bigg[ \sigma^2 + \frac{\sigma (\sigma^2-2)}{(\sigma^2-1)^{\frac{1}{2}}}\cosh^{-1} (\sigma)\Bigg] + {\cal O}(\epsilon)\; .
\label{Redel2}
\end{equation} 
This contribution emerges from the inclusion of radiation modes in the loop integrals and gives rise to half-integer-PN corrections to the deflection angle. 

Considering the full massive ${\cal N}=8$ result~\cite{DiVecchia:2021bdo}, we then noticed a simple relation between the contribution in eq.~\eqref{Redel2} and two terms appearing in the imaginary part of the same eikonal phase so that, in the full expression for $\delta_2^{(rr)}$, there are three terms that appear in the following combination:
\begin{equation}
 \left[
	1 + \frac{i}{\pi} 
	\left(
	- \frac{1}{\epsilon} + \log(\sigma^2-1)
	\right) 
	\right]  
	{\rm Re}\,2\delta_2^{(rr)}.
	\label{comb2}
\end{equation} 
The two imaginary contributions to $2 \delta_2^{(rr)}$ that appear in \eqref{comb2} are an IR-singular term, which captures the full contribution proportional to $\epsilon^{-1}$, and a $\log(\sigma^2-1)$ term, which captures the branch cuts starting at $\sigma=\pm1$.

Let us now examine whether this feature is to be regarded as an accident of the maximally supersymmetric theory or rather as a more general fact. 
As we shall discuss below and will explain in more detail in \cite{DiVecchia:2021bdo}, the precise combination of the two imaginary terms in the round bracket of \eqref{comb2} is dictated by the three-particle unitarity cut, where the phase space integration over the soft momentum of the massless quantum is responsible for the infrared singularity in ${\rm Im}\,2\delta_2$ (let us recall that ${\rm Im}\,2\delta_2$ contains just the inelastic contribution to the cut \cite{Amati:1990xe}).
Furthermore, using real-analyticity of the amplitude forces the $\log (\sigma^2 -1)$ to appear in $\delta_2$ as  $\log (1-\sigma^2) = \log (\sigma^2 -1) - i \pi$ yielding precisely the analytic structure of \eqref{comb2}. 
Combining these two observations, which are based purely on unitarity, analyticity and crossing symmetry,  we are led to conjecture the validity of \eqref{1.5} independently of the specific theory under consideration. 

As anticipated, this relation opens the way to a much simpler calculation of  RR effects since it trades the computation of $\operatorname{Re}2\delta_2^{(rr)}$ to that of the IR-divergent part of ${\rm Im}\,2\delta_2$. In the following sections we will carry out this calculation both for the supersymmetric case at hand, for pure gravity where we shall recover a recent result by Damour~\cite{Damour:2020tta}, and for the scalar-tensor theory of Jordan-Brans-Dicke.

For the purpose of computing the IR-divergent piece in ${\rm Im}\,2\delta_2$, one can focus on the leading ${\cal O}(k^{-1})$ term in the  soft expansion of the inelastic amplitudes given in Sect~\ref{softm}. This allows us to factor out, for each specific theory, the corresponding elastic amplitude. Next, and in this  order, one has to take the leading term in a small-$q$ expansion so as to get the sought-for classical contribution. In terms of the impact parameter $b$ which will be introduced in~\eqref{eq:ft}, the small-$q$ limit is equivalent to an expansion for large values of $b$. Since the soft factor is linear in $q$ (it goes to zero at zero scattering angle), and the tree amplitude has a $q^{-2}$ singularity, the result for the inelastic amplitude is  (modulo $\epsilon$ dependence) of $\mathcal O(b^{-1})$ and thus of the desired $\mathcal O(b^{-2})$ in ${\rm Im}\, 2\delta_2$. 

\section{Soft Amplitudes in $b$-space}
\label{softb}

We now start from the momentum space soft amplitudes given in Sect.~2 and go to impact parameter space using for a generic amplitude the notation
\begin{equation}\label{eq:ft}
	\tilde A(b) = \int \frac{d^{2-2\epsilon} q}{(2\pi)^{2-2\epsilon}}\frac{A(q)}{4m_1 m_2 \sqrt{\sigma^2-1}}\, e^{i b \cdot q}\;.
\end{equation}
We can now simply replace the factors of $q_j$ in the numerators of the various amplitudes $A_5$ by the derivative $- i \frac{\partial}{\partial b^j}$ and then perform the Fourier transform where the $q$-dependence appears only in $A_{tree}$.  Starting from the ${\cal N}=8$ elastic tree-level amplitude with $\phi=\frac{\pi}{2}$, given, up to analytic terms as $q^2\to0$, by
\begin{equation}
A_{tree} = 8 \pi \beta(\sigma) \frac{m_1m_2}{q^2} \,,
\qquad
\beta(\sigma) = 4 G m_1 m_2 \sigma^2\,,
\label{A0}
\end{equation}
the leading eikonal  takes the form
\begin{equation}
2 \delta_0 = -\beta(\sigma) \frac{\Gamma(1-\epsilon) (\pi b^2)^{\epsilon}}{2 \epsilon \hbar \sqrt{\sigma^2-1}} ~ \Rightarrow - i \frac{\partial}{\partial b^j} 2 \delta_0 = \frac{ i\,\Gamma(1-\epsilon)\, b^j (\pi b^2)^{\epsilon}}{ b^2 \hbar  \sqrt{\sigma^2-1}} \beta(\sigma)\;.
\label{delta0}
\end{equation}
As clear from~\eqref{2.1b}, one can move from $\mathcal N=8$ to the case of pure GR simply by replacing the prefactor $\beta(\sigma)$ by
\begin{equation}
  \label{eq:tAd}
  \beta^{GR}(\sigma) = 2  G m_1 m_2 (2\sigma^2-1)\;.
\end{equation}

We then obtain the following result for the classical part of the soft graviton and soft dilaton amplitudes in impact parameter space
  \begin{align}
  	\label{2.81l}
  	\begin{split}
  {\tilde{A}}_5^{\mu \nu} (\sigma, b, k) &\simeq {i  \frac{\kappa \beta(\sigma)(\pi b^2)^{\epsilon}}{b^2 \sqrt{\sigma^2-1}}} \\
  &\times \left[ (k b)\left(  \frac{\bar{p}_1^\mu \bar{p}_1^\nu}{(\bar{p}_1k)^2} - \frac{\bar{p}_2^\mu \bar{p}_2^\nu}{(\bar{p}_2k)^2}\right)- \frac{\bar{p}_1^\mu {b}^\nu+ \bar{p}_1^\nu {b}^\mu}{ (\bar{p}_1k) } + \frac{\bar{p}_2^\mu {b}^\nu+ \bar{p}_2^\nu {b}^\mu}{ (\bar{p}_2 k)  }  \right]\,,
  \end{split}
   \\
{\tilde{A}}_5^{dil} (\sigma, b, k) &\simeq - {i  \frac{\kappa \beta(\sigma)(\pi b^2)^{\epsilon}}{ \sqrt{2(\sigma^2-1)}}}  {\frac{(k b)}{b^2} } \left[ \frac{m_1^2 }{(\bar{p}_1k)^2} -  \frac{m_2^2}{(\bar{p}_2k)^2} \right] \,,
\label{2.82l}
\end{align}
where we approximated the factor of $\Gamma(1-\epsilon)$ in~\eqref{delta0} to 1 as we are interested in the $D\to 4$ case, but we continue to keep track of the dimensionful factor of $b^{2\epsilon}$.

Having obtained Eqs.~\eqref{2.81l}, \eqref{2.82l} with the appropriate normalization, we follow the same procedure to go over to $b$-space for the other fields relevant to the ${\cal N}=8$ analysis. For the two vectors we obtain
\begin{align}
  {\tilde{A}}_5^\mu &\simeq {i \sqrt{2} \frac{\kappa m_1 \beta(\sigma)(\pi b^2)^{\epsilon}}{ b^2 \sqrt{\sigma^2-1}} } \left[ \frac{ (k b) \bar{p}_1^\mu}{ (\bar{p}_1k)^2}- \frac{{b}^\mu}{ \bar{p}_1k}\right]\,, \\
   {\tilde{B}}_5^\mu &\simeq  {- i \sqrt{2} \frac{\kappa m_2 \beta(\sigma)(\pi b^2)^{\epsilon}}{ b^2 \sqrt{\sigma^2-1}} } \left[ \frac{ (k b)\bar{p}_2^\mu}{(\bar{p}_2k)^2}- \frac{{b}^\mu}{\bar{p}_2k}\right],
\label{2.10}
\end{align}
while for the two scalars we get
\begin{eqnarray}
{\tilde{A}}_5 \simeq{i \frac{\kappa m_1^2 \beta(\sigma)(\pi b^2)^{\epsilon}}{ b^2 \sqrt{\sigma^2-1}} } \frac{(k b)}{(\bar{p}_1k)^2} \,,\qquad
{\tilde{B}}_5 \simeq - {i \frac{\kappa m_2^2 \beta(\sigma)(\pi b^2)^{\epsilon}}{ b^2 \sqrt{\sigma^2-1}} } \frac{(k b)}{(\bar{p}_2 k)^2}\,.
\label{2.11}
\end{eqnarray}
Note that all our soft amplitudes are homogeneous functions of $\omega$ and $b$  of degree $-1$ and $-1 + 2 \epsilon$, respectively.

\section{IR Divergence of the 3PM Eikonal}
\label{probability}

Motivated by the discussion of Sect.~\ref{RRIR} and armed with the results of the Sect.~\ref{softb}, we now turn to the calculation of the infrared divergent part of ${\rm Im} \,2 \delta_2$ from the three-particle unitarity cut.
Indeed the unitarity convolution in momentum space diagonalizes in impact parameter space giving (see e.g. \cite{Amati:2007ak})
\begin{equation}
2 {\rm Im} \,2 \delta_2  = \sum_i \int \frac{d^{D-1} \vec k}{2|\vec{k}| (2\pi)^{D-1}} | \tilde{A}_{5i} |^2 \,,
\label{3.1}
\end{equation}
where the sum is over each massless state in the theory under consideration. For spin-one and spin-two particles this also includes a sum over helicities. Instead of separating different helicity contributions, we use the fact that all the $2 \to 3$ amplitudes we use are gauge invariant/transverse and simply insert the corresponding on shell Feynman and de Donder propagators, {\em i.e.} $\eta^{\mu\nu}$ for the vectors and $\frac{1}{2} \left(\eta^{\mu\rho} \eta^{\nu\sigma} + \eta^{\mu \sigma} \eta^{\nu \rho} -  \eta^{\mu\nu} \eta^{\rho\sigma} \right)$ for the graviton.

Equation \eqref{3.1} implies that $\beta^2(\sigma)$ always factors out of the integral over $\vec k$. In spherical coordinates the latter splits  into an integral over the modulus $|\vec{k}|$ and one over the angles defined by the following parametrisation of the vector $\vec{k}$:
\begin{equation}
\vec{k} = |\vec{k}| ( \sin \theta \cos \varphi,\, \sin \theta \sin \varphi,\, \cos \theta)\,,\qquad  
(k b) =  - |\vec{k}| b \sin \theta \cos \varphi\, ,
\label{3.1a}
\end{equation}
that implies
\begin{equation}
  \label{eq:pkang}
  (\bar{p}_1k) = |\vec{k}| (E_1 - \bar{p} \cos\theta)  \,, \qquad
  (\bar{p}_2k) = |\vec{k}| (E_2 + \bar{p} \cos\theta)\, ,
\end{equation}
where we have taken $b$ in \eqref{3.1a} along the $x$ axis. It is clear that the integral over $|\vec{k}| = \hbar \omega$ in~\eqref{3.1} factorises together with an $\epsilon$-dependent power of $b$ to give\footnote{We need to keep $D=4-2\epsilon$ only for the integral over $|\vec{k}|$ while the integration over the angular variables can be done for $\epsilon=0$, so that effectively $d^{D-1} \vec k = |\vec{k}|^{2-2\epsilon} d|\vec{k}|\, \sin \theta \,d \theta\, d \varphi$.}
\begin{equation}
2 {\rm Im} \,2 \delta_2  \sim  \int \frac{d \omega}{\omega} \omega^{- 2 \epsilon} (b^2)^{-1 +2 \epsilon} \sim (b^2)^{-1+ 3 \epsilon}  \int \frac{d \omega}{\omega} (\omega b)^{- 2 \epsilon}
\label{eq:of}
\end{equation}
where the factor $ (b^2)^{-1+ 3 \epsilon}$ is precisely the one expected (also on dimensional grounds) to appear in $\delta_2$. On the other hand, the integral over $\omega$ produces a $\frac{1}{\epsilon}$ divergence in the particular combination:
 \begin{equation}
 \int \frac{d \omega}{\omega} (\omega b)^{- 2 \epsilon} = - \frac{1}{2 \epsilon} (\,\overline{\omega b}\,)^{-2 \epsilon} =  -\frac{1}{2 \epsilon}  + \log \overline{\omega b}+ {\cal O}(\epsilon)
\label{comb}
\end{equation}
where $\overline{\omega b}$ is an appropriate upper limit on the classical dimensionless quantity $\omega b$.

To determine $\overline{\omega b}$ one can argue as follows.
By energy conservation:
\begin{equation}
\label{hom}
\hbar \omega = \Delta E_1 +  \Delta E_2
\end{equation}
where $\Delta E_i$ is the energy loss for the $i^{\rm th}$ particle.
On the other hand, in order for the spatial components of the momentum transfers $q_i = -(p_i + k_i)$ to provide  a classical contribution, they should be of order $\hbar/b \ll |\vec{p}_i|$. But then we can estimate \eqref {hom} by using (for on-shell particles): 
\begin{equation}
\label{onshell}
\Delta E_i \lesssim\frac{ |\vec{p}_i|}{E_i} |\Delta \vec{p}_i|   \qquad (i = 1,2)\,.
\end{equation}
Combining \eqref{hom} and \eqref{onshell} we arrive at 
\begin{equation}
\omega b  \lesssim \frac{ |\vec{p}_1|}{E_1}+ \frac{ |\vec{p}_2|}{E_2}\,.
\end{equation}
Using now the following (centre-of-mass) expressions,
\begin{equation}
    \label{eq:rws}
\begin{gathered}
  \bar{p} \; \simeq \; |\vec{p}\,| = \frac{m_1 m_ 2 \sqrt{\sigma^2-1}}{\sqrt{m_1^2 +m_2^2+2m_1 m_2 \sigma}}\;,\\
  E_1 =  m_1 \frac{m_1+ \sigma m_2}{\sqrt{m_1^2 +m_2^2+2 m_1 m_2 \sigma}}\;,\quad
  E_2 =  m_2 \frac{m_2+ \sigma m_1}{\sqrt{m_1^2 +m_2^2+2 m_1 m_2 \sigma}}\;,
\end{gathered}
\end{equation}
we find: 
\begin{equation}
\overline{\omega b} \sim  \sqrt{\sigma^2-1} \left( \frac{m_2}{m_1 + \sigma m_2} + \frac{m_1}{m_2 + \sigma m_1}  \right) = \sqrt{\sigma^2-1}( 1 + {\cal O}(\sigma -1))\;.
\label{baromb}
\end{equation}
Therefore, inserting this result in \eqref{comb} and using  the real-analyticity argument mentioned in Sect.~3,  precisely the combination appearing in \eqref{comb2} is indeed recovered. This is the essence of our argument for conjecturing \eqref{comb2} as a general connection between RR and soft limits. The rest of this section provides examples and non trivial tests of such a connection.
 
\subsection{Massive ${\cal{N}}=8$ Supergravity}

We evaluate separately the ${\cal O}(\epsilon^{-1})$ contribution to~\eqref{3.1} for each massless state: the graviton, the dilaton, two vectors and two scalars coupling to the particle of mass $m_1$ and other two vectors and two scalars  coupling to the particle of mass $m_2$. We first start from the dilaton contribution. By using~\eqref{2.82l} in~\eqref{3.1} we obtain 
\begin{equation}
  \label{eq:dp1}
  ({\rm Im} \,2 \delta_2)_{dil} \simeq \frac{\kappa^2 {\beta}^2(\sigma)}{4 b^2(\sigma^2-1)} \int  \frac{d |\vec{k}| |\vec{k}|^{-2\epsilon-1}  }{2 (2\pi)^{3}} \!\int_{-1}^1\!\!\!dx\,\pi (1-x^2)\!  \left[ \frac{m_1^2}{({E_1} -\bar{p} x)^2} -  \frac{m_2^2}{({E_2} +\bar{p} x)^2} \right]^2\!\!\!,
\end{equation}
where  $x=\cos\theta$. The extra factor of $\pi\sin^2\theta=\pi (1-x^2)  $ in the integrand follows from the integration over the angle $\varphi$. As already mentioned the integral over $|\vec{k}|$ factorises out of the whole integral and provides the sought for $\epsilon^{-1}$ factor. Finally, by using~\eqref{eq:rws}, we express everything in terms of $\sigma$ introduced in \eqref{2.1}. Then, using~\eqref{eq:rws} in~\eqref{eq:dp1} and performing the integral over $x$, we obtain
\begin{eqnarray}
  ({\rm Im} \,2 \delta_2)_{dil}   (\sigma, b) \simeq -\frac{1}{2 \epsilon } \frac{G {\beta}^2(\sigma)}{\pi \hbar b^2 (\sigma^2-1)^2}
  \left[ \frac{\sigma^2+2}{3} - \frac{\sigma}{ (\sigma^2-1)^{\frac{1}{2}} }
\cosh^{-1} (\sigma)  \right] .
\label{3.3}
\end{eqnarray}
Note that the final result depends on the masses only through $\sigma$ even if the integrand depends on $m_1, m_2$ and $\sigma$ separately. The term with the factor of $\cosh^{-1}(\sigma)$ emerges from the cross-product of the square in~\eqref{eq:dp1}, while the other terms yield only rational contributions in $\sigma$.

For the graviton's contribution,  using~\eqref{2.81l} in~\eqref{3.1}, we obtain %
\begin{align}
  ({\rm Im} \,& 2 \delta_2)_{gr}  (\sigma, b) \simeq - \frac{\kappa^2 {\beta}^2(\sigma)}{2 b^2 (\sigma^2-1)} \left(-\frac{1}{2 \epsilon} \,\frac{1}{2 (2\pi)^{3}} \right)
\pi  \int_{-1}^1\!\! dx  \nonumber \\
  \times & \Bigg\{4 \left[ \frac{m_1^2}{(E_1-\bar{p}x)^2} +  \frac{m_2^2}{(E_2+\bar{p} x)^2} - \frac{2m_1m_2 \sigma}{(E_1-\bar{p} x) (E_2 + \bar{p} x)}  \right] \\ \nonumber
& - \frac{1-x^2}{2}  \left[ \frac{m_1^4}{(E_1- \bar{p} x)^4} +  \frac{m_2^4}{(E_2+ \bar{p}x)^4} - \frac{2 m_1^2m_2^2 (2\sigma^2-1)}{(E_1 -\bar{p} x)^2 (E_2+\bar{p} x)^2}\right] \Bigg\}\,.
\end{align}
The integral over $x$ is again elementary\footnote{Surprisingly, it turns out to be the same as the integral appearing in Eq.~(4.4) of~\cite{Damour:2020tta} and thus reproduces exactly the function ${\cal I}$ in~(4.7) of that reference.}. 
 In terms of the variable $\sigma$ we obtain:
\begin{equation}
  ({\rm Im} \,2 \delta_2)_{gr} (\sigma, b) \simeq -\frac{1}{2 \epsilon } \frac{G {\beta}^2(\sigma)}{\pi \hbar b^2 (\sigma^2-1)^2}
  \left[ \frac{ 8 -5 \sigma^2}{3}  - \frac{\sigma(3-2\sigma^2)}{(\sigma^2-1)^{\frac{1}{2}}} \cosh^{-1} (\sigma) \right] .
\label{3.2}
\end{equation}

Following the same procedure for the contribution of the two vectors in~\eqref{2.5} we get
\begin{eqnarray}
  ({\rm Im} \,2 \delta_2)_{vec} (\sigma, b) \simeq -\frac{1}{2 \epsilon} \frac{G {\beta}^2(\sigma)}{\pi \hbar b^2 (\sigma^2-1)^2} \left[\frac{8}{3} (\sigma^2-1)\right]
\label{3.4}
\end{eqnarray}
and for the sum of the two scalars in~\eqref{2.6} we obtain
\begin{eqnarray}
  ({\rm Im} \,2 \delta_2)_{sca} (\sigma, b)\simeq -\frac{1}{2 \epsilon} \frac{G {\beta}^2(\sigma)}{\pi \hbar b^2 (\sigma^2-1)^2} \left[\frac{2}{3} (\sigma^2-1)\right].
\label{3.5}
\end{eqnarray}
In the last two types of contributions the soft particles are attached to the same massive state, so there are no terms in the integrand with the structure appearing in the cross term of~\eqref{eq:dp1} and hence no factors of $\cosh^{-1}(\sigma)$ in the final result. Also the graviton and the dilaton results contain contributions of this type corresponding to the terms in the integrands which depend only on $E_1$ or $E_2$. In the ${\cal N}=8$ setup these contributions cancel when summing over all soft particles. Notice also that the static limit $\sigma\to 1$ of~\eqref{3.4} and~\eqref{3.5} is qualitatively different from that of the full graviton and dilaton contributions as it starts one order earlier. Then thanks to~\eqref{1.5} also the leading term of the PN expansion of the ${\cal N}=8$ eikonal or deflection angle is due to the vectors and the scalars in~\eqref{2.3} and~\eqref{2.3b}.

By summing the contributions~\eqref{3.3}--\eqref{3.6}, we get the following result for the infrared divergent part of the three-particle discontinuity in ${\cal N}=8$ supergravity with $\phi=\frac{\pi}{2}$
\begin{eqnarray}
  \left({\rm Im} 2\delta_2\right) \simeq -\frac{G {\beta}^2(\sigma)}{\pi \hbar b^2 \epsilon (\sigma^2-1)^2} 
  \left[ \sigma^2 + \frac{\sigma (\sigma^2-2)}{(\sigma^2-1)^{\frac{1}{2}}} \cosh^{-1} (\sigma)  \right]   \;
\label{3.6}
\end{eqnarray}
and we can check that it is consistent with~\eqref{1.5} and~\eqref{Redel2}. 

Further checks of the relation~\eqref{1.5} could be performed by extending the same analysis to the case of $\cos \phi \ne 0$ or to supergravity theories with $0<\mathcal {N} < 8$.

\subsection{General Relativity and Jordan-Brans-Dicke Theory}

The calculation in pure GR follows exactly the same steps with only the contribution of the graviton and yields again the result in Eq.~\eqref{3.2}  just with the prefactor $({\beta}^{GR}(\sigma))^2$ in place of ${\beta}^2(\sigma)$. Then, assuming that Eq.~\eqref{1.5} is also valid in GR, we get 
\begin{equation}
  ({\rm Re} \,2 \delta^{(rr)}_2)_{GR} (\sigma, b) = \frac{G ({\beta}^{GR}(\sigma))^2}{2 \hbar b^2 (\sigma^2-1)^2} 
  \left[ \frac{ 8 -5 \sigma^2}{3}  - \frac{\sigma(3-2\sigma^2)}{(\sigma^2-1)^{\frac{1}{2}}} \cosh^{-1} (\sigma) \right]
\label{3.2b}
\end{equation}
and, from it, we obtain the deflection angle
\begin{equation}
  \label{eq:6.6}
(\chi^{(rr)}_3)_{GR} = - \frac{\hbar}{|\vec{p}|} \frac{\partial {\rm Re} 2 \delta^{(rr)}_2}{\partial b} =  \frac{G ({\beta}^{GR}(\sigma))^2}{|\vec{p}|  b^3 (\sigma^2-1)^2}
\left[ \frac{ 8 -5 \sigma^2}{3}  - \frac{\sigma(3-2\sigma^2)}{(\sigma^2-1)^{\frac{1}{2}}} \cosh^{-1} (\sigma) \right] 
\end{equation}
which reproduces the one given in Eq. (6.6) of~\cite{Damour:2020tta}. At the moment, the physical reason for this agreement is unclear.

The results obtained so far allow one to derive in a straightforward way the zero-frequency limit (ZFL) of the energy spectrum $\frac{dE^{rad}}{d\omega}$. Indeed,  the energy spectrum is just the integrand of~\eqref{3.1} for the graviton multiplied by an extra factor of $ \hbar \omega$ (see also~\cite{Goldberger:2016iau}) so that,
\begin{equation}
  \label{eq:EradG3}
  E^{\rm rad} = \int \frac{d^{D-1} k}{2 (2\pi)^{D-1}} \tilde{A}^{*\, \mu\nu}_{5}\left(\eta_{\mu\rho} \eta_{\nu\sigma} - \frac{1}{2} \eta_{\mu\nu} \eta_{\rho\sigma}\right) \tilde{A}_5^{\rho\sigma} \equiv \int_0^\infty\!\!\! d\omega \,\frac{d E^{rad}}{d\omega}\;.
\end{equation}
Since we computed only the $k\to 0$ limit of this integrand, we can reliably extract just the ZFL
\begin{equation}
  \frac{d E^{rad}}{d\omega} ( \omega\to 0) =   \lim_{\epsilon\to 0} \left[-4 \hbar \epsilon ({\rm Im} 2\delta_2)\right] \,.
\label{3.7}
\end{equation}
In the case of GR we can use~\eqref{3.2} with $({\beta}^{GR}(\sigma))^2$ in place of ${\beta}^2(\sigma)$ and reproduce Eq.~(2.11) of~\cite{Kovacs:1978eu} (taken from~\cite{Ruffini:1970sp}) by taking the static limit $\sigma \to 1$
\begin{equation}
\frac{dE}{d \omega} ( \omega \to 0)  = \frac{32 G^3 m_1^2 m_2^2}{5\pi b^2}\;.
\label{3.12}
\end{equation}

Our result \eqref{3.7} should hold true\footnote{T. Damour kindly informed us that he has carried out the explicit check.}  at all values of $\sigma$, extending Smarr's original result \cite{Smarr:1977fy} to arbitrary kinematics  (see~\cite{Kovacs:1978eu}). Possibly, our approach can be  extended to compute the energy spectrum to sub and sub-sub leading order in $\omega$ and to reproduce, in particular cases, the results of~\cite{Sahoo:2018lxl}, \cite{Addazi:2019mjh} and~\cite{Saha:2019tub}. 

On the other hand, our method looks inadequate to deal with the full spectrum and with the total energy loss\footnote{Such a calculation has been recently tackled  by a different approach in \cite{Herrmann:2021lqe}.}. For instance, extrapolating the ZFL result \eqref{3.7} to the upper limit given in \eqref{baromb}  would reproduce, at large $\sigma$, the qualitative behaviour of Eq. (5.10) of \cite{Kovacs:1978eu}. But, as anticipated to be the case in  \cite{Kovacs:1978eu}, and discussed in \cite{Gruzinov:2014moa} and \cite{Ciafaloni:2018uwe}, such a result needs to be amended, as in the ultra-relativistic/massless limit, at fixed $G$, it would violate energy conservation. 

Our connection between RR and soft limits readily applies to Jordan-Brans-Dicke (JBD) scalar-tensor  theory.
The coupling of the massless scalar to massive particles is very much like that of the dilaton except for a rescaling of the coupling by a function of the JBD parameter $\omega_J$ 
(the coefficient of the JBD kinetic term):
\begin{equation}
g_{JBD} = \frac{1}{\sqrt{2 \omega_J + 3}}\, g_{dil}
\label{JBDcoup}\,.
\end{equation}
The string dilaton case is recovered for $\omega_J = -1$.
It is then straightforward to calculate the RR in JBD theory. It amounts to inserting in~\eqref{3.2} and in~\eqref{3.3} the JBD  $\beta(\sigma)$ factor,
\begin{equation}
\beta^{JBD}(\sigma)  =  4G m_1 m_2 \left(\sigma^2- \frac{\omega_J +1}{2 \omega_J + 3}\right),
\label{JBDbeta}
\end{equation}
 and to further multiplying the dilaton's contribution of \eqref{3.3} by a factor $(2 \omega_J + 3)^{-2}$. Thus the contribution to the radiation reaction part of the eikonal from the JBD scalar reads
 \begin{equation}
   \frac{G {(\beta^{JBD})}^2 (2 \omega_J + 3)^{-2}}{2 \hbar b^2 (\sigma^2-1)^2}
  \left[ \frac{\sigma^2+2}{3} - \frac{\sigma}{ (\sigma^2-1)^{\frac{1}{2}} }
\cosh^{-1} (\sigma)  \right]  .
\label{JBDdelta}
 \end{equation}
 In the limit $\omega_J \to \infty$, this result vanishes leaving just the contribution of the graviton and thus reproducing the GR result.   Since the present lower limit on $\omega_J$ is about $4\times 10^4$ the effect is unfortunately unobservable.

\subsection*{Acknowledgements} 
We thank Enrico Herrmann, Julio Parra-Martinez, Michael Ruf and Mao Zeng for sharing with us a first draft of their paper \cite{Herrmann:2021lqe} and for useful comments on ours.
We also thank Zvi Bern, Emil Bjerrum-Bohr, Poul Henrik Damgaard, Thibault Damour,  Henrik Johansson, Rafael Porto and Ashoke Sen for valuable observations on a preliminary version of this letter.
 The research of RR is partially supported by the UK Science and Technology Facilities Council (STFC) Consolidated Grant ST/P000754/1 ``String theory, gauge theory and duality''. The research of CH (PDV) is fully (partially) supported by the Knut and Alice Wallenberg Foundation under grant KAW 2018.0116.

\providecommand{\href}[2]{#2}\begingroup\raggedright\endgroup

\end{document}